\documentclass[aps,prl,preprint,groupedaddress,longbibliography]{revtex4-1}

\usepackage{amsmath}
\usepackage{graphicx}
\usepackage{hyperref}

\newcommand{\ket}[1]{\left| #1 \right\rangle}
\newcommand{\bra}[1]{\left\langle #1 \right|}
\newcommand{\braket}[2]{\left\langle #1 \middle| #2 \right\rangle}
\DeclareMathOperator{\sgn}{sgn}
\DeclareMathOperator{\id}{I}
\DeclareMathOperator{\Ex}{\mathcal{E}_\textit{x}}
\DeclareMathOperator{\tr}{tr}
\DeclareMathOperator{\odd}{odd}
\DeclareMathOperator{\erf}{Erf}

\begin{document}

\title{What is measured when a qubit measurement is performed on a multi-qubit chip?}

\author{Joel C. Pommerening}
\author{David P. DiVincenzo}

\affiliation{Institute for Quantum Information, RWTH Aachen University, D-52056 Aachen, Germany}
\affiliation{Peter Gr\"unberg Institute, Theoretical Nanoelectronics, Forschungszentrum J\"ulich, D-52425 J\"ulich, Germany}
\affiliation{J\"ulich-Aachen Research Alliance (JARA), Fundamentals of Future Information Technologies, D-52425 J\"ulich, Germany}

\date{\today}

\begin{abstract}
We study how single-qubit dispersive readout works alongside two qubit coupling. To make calculations analytically tractable, we use a simplified model which retains core characteristics of but is discretised compared to dispersive homodyne detection. We show how measurement speed and power determine what information about the qubit(s) is accessed. Specifically we find the basis the measurement is closest to projecting onto. Compared to the basis gates are applied in, this measurement basis is modified by the presence of photons in the readout resonator.
\end{abstract}

\maketitle

\section{Introduction}
Dispersive readout \cite{Blais04}  is a well established measurement technique in the toolbox of circuit quantum electrodynamics. A fidelity of 99.2\% has been achieved experimentally for single-qubit readout \cite{Wallraff17}, and high-fidelity multiplexed readout has been demonstrated as well, e.g. in Ref.~\cite{Wallraff18} for five qubits with an average accuracy of 97\%. For near-term applications this is sufficient \cite{Wendin17}. Beyond that, when targeting more complex circuits, in particular those involving feedback from intermediate measurements, even lower error rates can be expected to become necessary. It is therefore important to identify potential error sources and anticipate bottlenecks.

Here we study how single-qubit dispersive measurement interfaces with networks of coupled qubits. Specifically we address the question, what exactly is being measured? This work serves to improve our fundamental understanding of the process, and shows how disregarding the effects of qubit-qubit coupling leads to new errors. These are especially relevant for quantum circuit operations that do not terminate after one measurement, since not only the distribution of outcomes but also the state after measurement are affected.
This problem has been addressed before with different methods \cite{Khezri15}, here we arrive at some of the same conclusions but also present new observations. Similar questions have also been studied in a different measurement setup \cite{Ashhab09PRA,Ashhab09NJP}.

Qubit coupling is necessary for facilitating two-qubit gates, but should be ``turned off'' otherwise. One way to do this, initially demonstrated in \cite{deGroot10}, is by keeping the qubits well separated in frequency, and having a fixed coupling---which is small compared to their frequency detuning---that is then activated by applying a cross-resonant drive. In this case the detuning of qubit frequencies must not be so large that it unduly slows down two-qubit gates but also not so small as to induce unwanted interaction. This middle-ground parameter regime is particularly interesting to us as it is most likely here that qubit-qubit interactions have a significant effect on single-qubit operations. Other coupling schemes can involve tuning either the qubits into resonance or a coupler on and off. Ref.~\cite{Gambetta17} gives a general overview of the different kinds of two-qubit gates for superconducting qubits and their respective merits and challenges. But in this paper we will focus on fixed-frequency qubits with constant coupling.

Acknowledging the influence of qubit-qubit interaction here means that the measurement does not commute with the system Hamiltonian and thus is not a priori quantum nondemolition (QND). We can still look for the best QND approximation of the measurement, by which we mean: Which basis is the measurement closest to (potentially noisily) projecting onto? In terms of the original physical qubits, this basis will include slightly entangled states. Crucially, we will present evidence that this measurement basis is not the same as the basis in which gates are applied. Interpreting the measurement as a projection on this new basis reveals additional information about the system state compared to treating it as a measurement in the original basis with a random error.

We use a model based on Ref.~\cite{Nigg13} which retains core characteristics of a dispersive readout but can be evaluated essentially analytically without invoking trajectory theory \cite{Blais04}. This makes the calculations tractable in a very wide range of parameters, including the experimentally relevant ones, and allows for straightforward interpretation. Thus we can develop and validate intuition about how measurement speed (relative to system dynamics) and distinguishing power influence precisely what information about the qubit(s) is accessed.

We demonstrate the concept on the smallest multiqubit network, i.e. two coupled qubits. We briefly review the level of modelling, then formalise the measurement model, and set up our expectations and method for analysing it, before presenting the results.

\section{Model}
 \begin{figure}
 \includegraphics[width=\columnwidth]{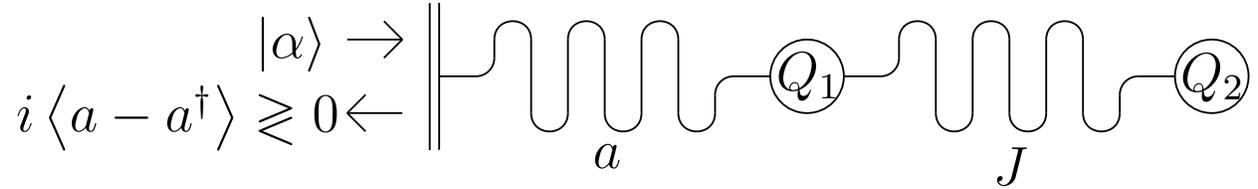}
 \caption{\label{fig1}Two qubits dispersively coupled to two resonators. Measurement consists of injecting a coherent state $\ket\alpha$ ($\alpha$ real) and detecting whether this state, after evolution in the first resonator, acquires a positive or a negative phase.}
 \end{figure}
The simplest network in which to observe the effect of qubit coupling is composed of two qubits, as shown in Fig.~\ref{fig1}. There are several ways to engineer a fixed coupling between qubits. Here we show a bus resonator that can be eliminated with a Schrieffer-Wolff transformation in favour of an effective direct interaction $J$, assuming the coupling is weak compared to the qubit-resonator detunings \cite{Richer13,Gambetta13}. Direct (e.g.  capacitive) coupling is also common. A second resonator is dispersively coupled for readout of qubit 1. Transforming to the rotating frame for the readout resonator, and ignoring the Lamb shift, we are left with only the qubit ac-Stark shift $\chi$, so our starting Hamiltonian is ($\hbar=1$) 
\begin{equation}\label{ham}
H=-\frac{\omega_1}{2}Z_1-\frac{\omega_2}{2}Z_2+\frac{J}{2}\left(X_1 X_2+Y_1 Y_2\right)+\chi Z_1 a^\dagger a .
\end{equation}
Here $X_i$, $Y_i$, $Z_i$ are the Pauli operators on qubit $i$, $a^{(\dagger)}$ annihilation (creation) operators on the readout resonator, $\omega_i$ the qubit frequencies, and $J$ is the effective qubit-qubit coupling, $i=1,2$.
The two resonators are assumed to be well separated in frequency, so we have neglected qubit-induced resonator-resonator coupling.
This is a fairly standard minimal model \cite{Blais04,Gambetta13,Magesan20}, and also the same one as studied in Ref.~\cite{Khezri15}.
This Hamiltonian is diagonal in the Fock basis of the readout resonator, and can easily be completely diagonalised by a resonator-occupation dependent rotation in the $\ket{01}$, $\ket{10}$ two-qubit subspace.
Two-qubit states are ordered as $\ket{ij}=\ket i_1\otimes\ket j_2$, with subscripts distinguishing qubits $1$ and $2$ where necessary.
The eigenstates are product states of the two-qubit states $\ket{\psi_n^k}$ and resonator Fock states $\ket n$ with eigenenergies $E_n^k$, given here for later reference,
\begin{widetext}
\begin{eqnarray}
\!\!\!\!\left\{\ket{\psi_n^k}\right\}_{k=1,..4}&=&\left\{\ket{00},\,\cos\gamma_n\ket{01}+\sin\gamma_n\ket{10},\,-\sin\gamma_n\ket{01}+\cos\gamma_n\ket{10},\,\ket{11}\right\}\label{eigenstates}\\
\left\{E_n^k\right\}_{k=1,..4}&=&\left\{-\frac{\omega_1+\omega_2}{2}+\chi n,\,\sgn\left(\delta_n\right)\sqrt{\delta_n^2+J^2},\,-\sgn\left(\delta_n\right)\sqrt{\delta_n^2+J^2},\,\frac{\omega_1+\omega_2}{2}-\chi n \right\}\,\,\,\,\,\,\,\,
\end{eqnarray}
\end{widetext}
with $\delta_n=(\omega_2-\omega_1)/2+\chi n$ and
\begin{eqnarray}
\cos\gamma_n&=&\frac{1}{\sqrt2}\sqrt{1+\frac{\left|\delta_n\right|}{\sqrt{\delta_n^2+J^2}}},\\
\sin\gamma_n&=&\frac{\sgn\left(J\delta_n\right)}{\sqrt2}\sqrt{1-\frac{\left|\delta_n\right|}{\sqrt{\delta_n^2+J^2}}}
\end{eqnarray}
so that
\begin{equation}
H=\sum_{n=0}^\infty\sum_{k=1}^4 E_n^k \ket{\psi_n^k}\bra{\psi_n^k}\otimes\ket{n}\bra{n}.
\end{equation}
Here the sign functions are added such that the diagonalising rotation becomes the identity in the limit of $J\to0$ with the convention
\begin{equation}
\sgn\left(x\right)=\begin{cases}\phantom{-}1, & x\ge0, \\ -1, & x<0. \end{cases}
\end{equation} 

We choose to analyse an alternative measurement protocol that still shares many characteristics of the standard homodyne measurement, but replaces the stochastic with unitary evolution followed by a single ideal measurement, thus becoming much simpler to treat theoretically. The constant drive of the readout resonator is replaced by initialisation to a coherent state. We then wait until the pointer states are maximally separated in phase space before reading them out in one step  thus discretising the measurement instead of continuously acquiring incremental information on the qubit state.

This scheme is based on Ref.~\cite{Nigg13} with two readout cavities, one with low quality factor and the other with high. The former implements ``instantaneous'' initialisation and readout of the latter and is not explicitly simulated. We will call this the Nigg-Girvin measurement in the following.

Algorithmically the measurement can be described as follows:
\begin{enumerate}
\item Initialise cavity to coherent state $\ket\alpha$ (assume $\alpha$ real and positive for concreteness).
\item Let cavity interact with qubit(s) for time-interval $t_m=\pi/2|\chi|$, as described by time evolution operator $U(t)=\exp(-iHt)$.
\item Readout cavity by POVM with elements \cite[Sec.~9-5]{Peres93}
\begin{equation}
E_\pm=\frac1\pi\int_{\Omega_\pm}\mathrm{d}^2\beta\, \ket\beta\bra\beta, \qquad E_++E_-=\id
\end{equation}
that are integrals over coherent states in the lower ($\Omega_{\sgn\chi}$) and upper halfplane ($\Omega_{-\sgn\chi}$), such that the unnormalised state after measurement is
\begin{equation}
\frac1\pi\int_{\Omega_\pm}\mathrm{d}^2\beta\, \ket\beta\bra\beta\rho\ket\beta\bra\beta
\end{equation}
with $\rho$ the state of the system just before the cavity measurement.
\item Trace out the cavity.
\end{enumerate}
We add the last step since we are primarily interested in a superoperator on the qubit Hilbert space only. Then this superoperator that describes the action of the measurement with result $x\in\{\pm\}$ on an initial two qubit state $\rho$ is
\begin{eqnarray}
\!\!\!\!\!\!\!\!\!\!\!\Ex\left(\rho\right)&=&\tr_\text{res}E_x U\!\left(t_m\right)\rho\otimes\ket\alpha\bra\alpha U^\dagger\!\left(t_m\right)\\
&=&\sum_{n,m=0}^\infty g_x(m,n)\bra n U\!\left(t_m\right)\ket n \rho\bra m U^\dagger\!\left(t_m\right) \ket m
\end{eqnarray}
where we evaluate the integrals in phase space to express $E_x$ in terms of the resonator Fock basis and get
\begin{eqnarray}
&&g_x(m,n)=\bra m E_x \ket n \braket n \alpha \braket \alpha m\\
&&=e^{-\alpha^2}\left(\frac{\alpha^{2n}}{2n!}\delta_{nm}-\frac{ix}{\pi}\frac{\alpha^{n+m} \Gamma\!\left(\frac{n+m}{2}+1\right)}{n! m! (m-n)}\odd (m-n)\right)\nonumber
\end{eqnarray}
with
\begin{equation}
\odd(n)=\begin{cases}1, &n\text{ is an odd integer,}\\0, &\text{else.}\end{cases}
\end{equation}
Note that $g_x(m,n)$ is peaked around $n,m=\alpha^2$ and falls off fast enough for large $n$, $m$ that we can generally evaluate the double sum by truncating it after some $n_\text{max}\gg\alpha^2$. With this approximation the superoperator can be evaluated completely analytically.

In the limit of $J\to0$, $\alpha\to\infty$ this would be a perfect projective measurement of qubit 1
\begin{equation}
\Ex^\text{ideal}(\rho)=\begin{cases}\ket{0}\bra{0}_1\rho\ket{0}\bra{0}_1, &x=+,\\
\ket{1}\bra{1}_1\rho\ket{1}\bra{1}_1, &x=-.\end{cases}
\end{equation}
However, we are not in the rotating frame of the qubits so we have to account for the time-evolution that has also taken place on the qubit subspace
\begin{equation}\label{ideal-meas}
\lim_{\substack{J\to0 \\ \alpha\to\infty}}\Ex(\rho)=\Ex^\text{ideal}\left(U_0\rho U_0^\dagger\right)
\end{equation}
with
\begin{equation}
U_0=\exp\left(\frac{it_m}{2}\left(\omega_1 Z_1+\omega_2 Z_2\right)\right).
\end{equation}
If we relax the $\alpha\to\infty$ limit, it is known that we get a single-qubit measurement with a finite signal-to-noise ratio (SNR) \cite{Gambetta07}
\begin{equation}\label{SNR-meas}
\lim_{J\to0}\Ex(\rho)=\Ex^\text{SNR}\left(U_0\rho U_0^\dagger\right),
\end{equation}
which we can e.g. characterise with a chi matrix as
\begin{eqnarray}
\Ex^\text{SNR}(\rho)&=&\sum_{i,j=0}^1\chi_{ij}^{x}\ket i \bra i_1 \rho \ket j \bra j_1 \\
\chi^x &=&\frac12 \begin{pmatrix}
1+x \erf\alpha & e^{-2\alpha^2} \\
e^{-2\alpha^2} & 1-x \erf\alpha\label{chimatrix}
\end{pmatrix}.
\end{eqnarray}
This way we include the inherent measurement error due to the finite overlap of coherent pointer states. Both of these idealised single-qubit models can serve as reference against which to compare our Nigg-Girvin measurement. Equation~(\ref{ideal-meas}) can be used to quantify the imperfection of the Nigg-Girvin measurement protocol, while Eq.~(\ref{SNR-meas}) is fine-tuned to isolate the effect of the coupling. In practice the difference between the two is very small for reasonably large $\alpha$.

\section{Measurement basis candidates}
In the ideal scenario for quantum computation, we like to think of single- and multi-qubit operations as independent building blocks of larger circuits that are supposed to work the same way when assembled into multi-qubit networks as these building blocks do in isolation. And it seems they mostly do, because experimental parameters are such that the rotating wave approximation (RWA) is applicable. RWAs are ubiquitous in the study of superconducting qubit systems, with the tacit understanding that the quality of a RWA depends strongly on the choice of a rotating frame.
One of the main questions we want to address in this work is: In which frame does the RWA, that turns our measurement model into a QND single-qubit operation, work best?

A frame is characterised by a qubit basis (the potential measurement basis) and corresponding frequencies. In this section we identify plausible basis candidates which form the starting point of our further analysis.

Given the Hamiltonian~(\ref{ham}), the simplest option is neglecting the $J$-coupling in a RWA, since typically $|J|\ll|\delta_n|$. The remaining terms in $H$ commute and can implement a measurement of $Z_1$. We call the common eigenbasis of $Z_1$, $Z_2$ the bare basis, namely the states
\begin{equation}\label{bare-basis}
\ket{00},\,\ket{01},\,\ket{10},\,\ket{11}
\end{equation}
with $Z_i=\ket0\bra0_i-\ket1\bra1_i$, $i=1,2$.

Outside of measurement (i.e. when the readout resonator is not occupied), the Hamiltonian is diagonalised by what we call the dressed basis, $\ket{\psi_0^{1,..4}}$ in Eq.~(\ref{eigenstates}), which we denote as
\begin{eqnarray}
\ket{\tilde0\tilde0}&=&\ket{00},\quad\phantom{-}\ket{\tilde0\tilde1}=\cos\gamma_0\ket{01}+\sin\gamma_0\ket{10}, \nonumber \\ \ket{\tilde1\tilde0}&=&-\sin\gamma_0\ket{01}+\cos\gamma_0\ket{10},\quad\ket{\tilde1\tilde1}=\ket{11}.\label{dressed-basis}
\end{eqnarray}
The two bases introduced so far were also considered in Ref.~\cite{Khezri15}, where dressed states are labeled $\ket{\overline{00}}$, $\ket{\overline{01}}$ etc., cf. also Ref.~\cite{Galiautdinov12}.
The Hamiltonian in the dressed basis has a similar form as in the bare basis, except the off-diagonal term becomes $n$-dependent
\begin{eqnarray}
H&=&\frac12\left(-\frac{\omega_1+\omega_2}{2}+\chi a^\dagger a\right)\left(\tilde Z_1+\tilde Z_2\right) \nonumber\\
&+&\frac12\left(\sgn\left(\delta_0\right)\sqrt{\delta_0^2+J^2}+\frac{\chi \left|\delta_0\right|}{\sqrt{\delta_0^2+J^2}} a^\dagger a\right)\left(\tilde Z_1-\tilde Z_2\right) \nonumber\\
&-&\frac{J\chi\sgn\left(\delta_0\right)}{2\sqrt{\delta_0^2+J^2}} a^\dagger a \left(\tilde X_1 \tilde X_2+\tilde Y_1 \tilde Y_2\right),\label{h-dressed}
\end{eqnarray}
with $\tilde X_i$, $\tilde Y_i$, $\tilde Z_i$, $i=1,2$, the Pauli operators in the dressed basis, e.g. $\tilde{Z}_1=\ket{\tilde0\tilde0}\bra{\tilde0\tilde0}+\ket{\tilde0\tilde1}\bra{\tilde0\tilde1}-\ket{\tilde1\tilde0}\bra{\tilde1\tilde0}-\ket{\tilde1\tilde1}\bra{\tilde1\tilde1}$.

In direct analogy to Eq.~(\ref{ideal-meas}), we can define an ideal dressed basis measurement as time-evolution with $H$ as in Eq.~(\ref{h-dressed}), with $a^\dagger a$ replaced by 0, followed by a projection on $\ket{\tilde0}_1$ or $\ket{\tilde1}_1$
\begin{eqnarray}
\ket{\tilde0}\bra{\tilde0}_1&=&\ket{\tilde0\tilde0}\bra{\tilde0\tilde0}+\ket{\tilde0\tilde1}\bra{\tilde0\tilde1}, \\
\ket{\tilde1}\bra{\tilde1}_1&=&\ket{\tilde1\tilde0}\bra{\tilde1\tilde0}+\ket{\tilde1\tilde1}\bra{\tilde1\tilde1}.
\end{eqnarray}
Similarly we can model an imperfect non-interacting  measurement with the same chi matrix as in Eq.~(\ref{chimatrix}), replacing the projection on the bare basis by projection on the dressed basis.

Dropping the off-diagonal part of Eq.~(\ref{h-dressed}) in a RWA is a good approximation if
\begin{equation}\label{RWA-cond}
\left\lVert\frac{J}{\delta_0+\chi a^\dagger a}\right\rVert \left\lVert\frac{\chi a^\dagger a}{\delta_0+\frac{J^2}{\delta_0+\chi a^\dagger a}}\right\rVert\ll1.
\end{equation}
The first factor is the same one that the bare basis RWA is conditional upon. Thus whether the second factor is less or greater than 1 determines if the dressed or bare basis is more suitable for the RWA. They are equally good (or bad) if $\lVert\chi a^\dagger a\rVert=\sqrt{\delta_0^2+J^2}$. Since Eq.~(\ref{RWA-cond}) depends on the population of the readout resonator, unless it is in a Fock state, we cannot make a definitive statement based solely on this simple comparison. Yet as we will see below, replacing $a^\dagger a$ by its expectation value does produce serviceable estimates, e.g. we will be seeing a crossover around $\pm\chi_c=\pm\sqrt{\delta_0^2+J^2}/\alpha^2$.

For typical parameters $|\chi|,\,|J|\ll|\delta_0|$ and small resonator occupation, Eq.~(\ref{RWA-cond}) suggest that the dressed basis provides a better approximation than the bare basis. Physically this means that if the speed ($1/\chi$) at which information ($\left\langle a-a^\dagger\right\rangle$) is acquired is slow compared to the system dynamics ($1/\delta_0$), the system can undergo many oscillations between bare basis states the measurement is attempting to project on, whereas the eigenstates are approximately stable. If instead the measurement were very fast, $|\chi|\gg|\delta_0|$, the $Z_1 a^\dagger a$ measurement Hamiltonian could achieve a projection on the bare basis before it was disturbed. These observations are in line with our expectation that the eigenbasis of the idling system is perhaps the most natural candidate for the qubit basis. But will the measurement projection be somewhat different still?

Bare and dressed bases are both special cases of Eq.~(\ref{eigenstates}) in the limits of $n\to\infty$ and $n=0$ respectively. This can be naturally extended to a discrete sequence of bases by including all the $n$, $\gamma_n$ in between and further to a continuous set of bases indexed by $\gamma$ corresponding to some real $n(\gamma)>0$. Here $n(\gamma)$ is defined such that the Hamiltonian is diagonal in the basis rotated by angle $\gamma$ when $a^\dagger a$ in Eq.~(\ref{ham}) is replaced by
\begin{equation}
n(\gamma)=\frac{\omega_1-\omega_2+2J\cot(2\gamma)}{2\chi}.
\end{equation}
For each basis from this continuum an ideal measurement can be constructed in the same way as for the dressed basis, except with $n(\gamma)$ instead of $n=0$. Taking our previous thoughts to their logical conclusion, we  are especially curious about $n$ equal to the expectation value of $a^\dagger a$. Then the $\gamma$-dressed basis looks the same way as in Eq.~(\ref{dressed-basis}), only with a different angle
\begin{equation}\label{na2-basis}
\ket{\tilde0\tilde1}=\cos\gamma_{n\to\alpha^2} \ket{01}+\sin\gamma_{n\to\alpha^2}\ket{10}
\end{equation}
etc. with
\begin{equation}
\tan2\gamma_{n\to\alpha^2}=\frac{J}{\delta_0+\chi\alpha^2}.
\end{equation}

Note that so far we make reference only to the Hamiltonian, and not to a specific measurement scheme.

\section{Diamond norm}
Now that we have discussed the different model measurements and how to represent them as superoperators, we need an appropriate metric to compare them. Since the two outcomes are symmetric in our model, it suffices to consider one.

While fidelity is a popular measure for comparing states or unitary gates, we now want to compare trace-decreasing, completely positive superoperators. In the following, let $A$ be a linear operator on a $d$-dimensional Hilbertspace with basis $\left\{\ket i\right\}$, $\mathcal{E}$ a superoperator acting on $A$, and
\begin{equation}
J(\mathcal{E})=\sum_{i,j}\mathcal{E}\left(\ket i \bra j \right)\otimes \ket i \bra j
\end{equation}
its Choi-Jamio{\l}kowski representation.
The trace (or 1-) norm is a straightforward operator norm that induces a superoperator norm by maximising over inputs
\begin{equation}
\left\lVert A\right\rVert_1=\tr\left(\sqrt{A^\dagger A}\right)\qquad\left\lVert\mathcal{E}\right\rVert_1=\max_{\lVert A\rVert_1\le 1} \left\lVert \mathcal{E}(A) \right\rVert_1.
\end{equation}
But since we are interested in the errors arising from performing single-qubit operations on networks of qubits, we want a norm stable under taking the tensor-product with identity. This leads us to the diamond norm, which gives a worst-case error rate \cite{Kitaev98}
\begin{equation}
\lVert\mathcal{E}\rVert_\diamond=\left\lVert\mathcal{E}\otimes\id_d\right\rVert_1.
\end{equation}

It can be efficiently computed using semidefinite programming (SDP) \cite{Watrous09,Watrous12}. The trace norm of the Choi-Jamio{\l}kowski map provides a bound on the diamond norm, which may also be used as a quick alternative to get an idea of the behaviour of a certain parameter set \cite[Sec. 3.4]{Watrous18}
\begin{equation}\label{bounds}
\frac1d\left\lVert J(\mathcal{E}) \right\rVert_1 \le \left\Vert \mathcal{E} \right \Vert_\diamond \le \left\Vert J(\mathcal{E}) \right\Vert_1 .
\end{equation}

\section{Results and discussion}
\begin{figure}
 \includegraphics[width=\columnwidth]{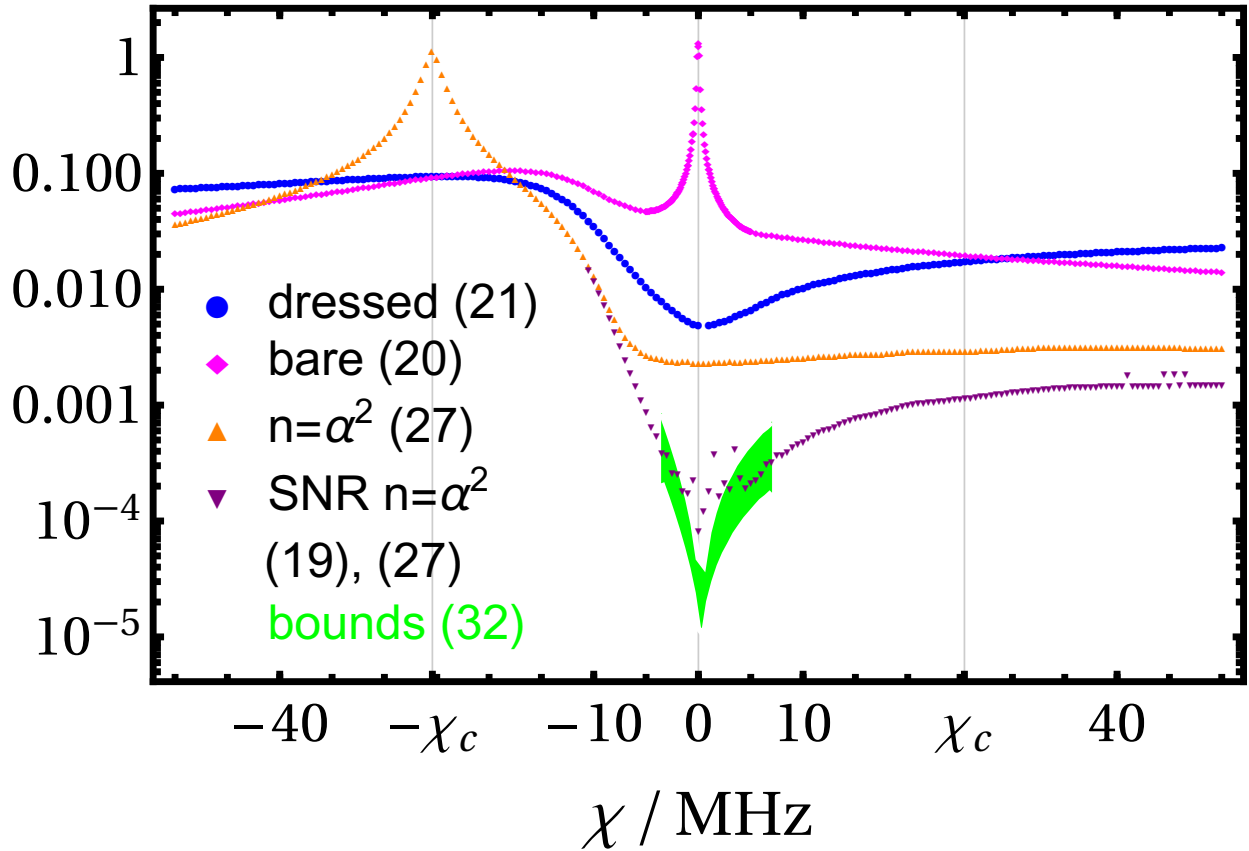}
 \caption{\label{fig2}Difference of Nigg-Girvin measurement superoperator to various idealised models, indexed by $i$ cf. the legend and equation numbers therein, as measured by the diamond norm $||\mathcal{E}-\mathcal{E}^i||_\diamond$ . The (realistic) parameters are $\delta_0=102\,$MHz, $J=3.8\,$MHz, $\alpha=2$ with $n_{\text{max}}=40$. The finite SNR version is shown where it visibly deviates from the idealised measurement; when it gets too small for the SDP-solver to handle, the bounds from Eq.~(\ref{bounds}) are shown in green.}
 \end{figure}

First let us examine the previously developed intuitive picture that slow (fast) measurements project onto the dressed (bare) basis. Figure~\ref{fig2} shows the deviation of the Nigg-Girvin measurement from an ideal measurement in the dressed or bare basis against $\chi$, to which measurement times $t_m=\pi/2|\chi|$ are inversely proportional.
The general behaviour in Fig.~\ref{fig2} is in line with our predictions, showing that for small $|\chi|$ the dressed basis approximation is very accurate, while for increasing $|\chi|$ the bare basis description improves and finally surpasses the dressed basis. This is in agreement with the conclusions drawn in Ref.~\cite{Khezri15}. The simple arguments following Eq.~(\ref{RWA-cond}) provide a reasonable order-of-magnitude estimate for the crossover ($\chi_c$) from bare to dressed basis behaviour, while slightly underestimating the crossing point.

Current experimental target parameters for $\chi$ in the single MHz regime \cite{Wallraff17} fall near the very centre of Fig.~\ref{fig2} where the dressed basis provides a better approximation than the bare basis measurement. The eventual crossover into bare basis behaviour is thus more of a mathematical rather than practical observation at this point.

\begin{figure}
 \includegraphics[width=\columnwidth]{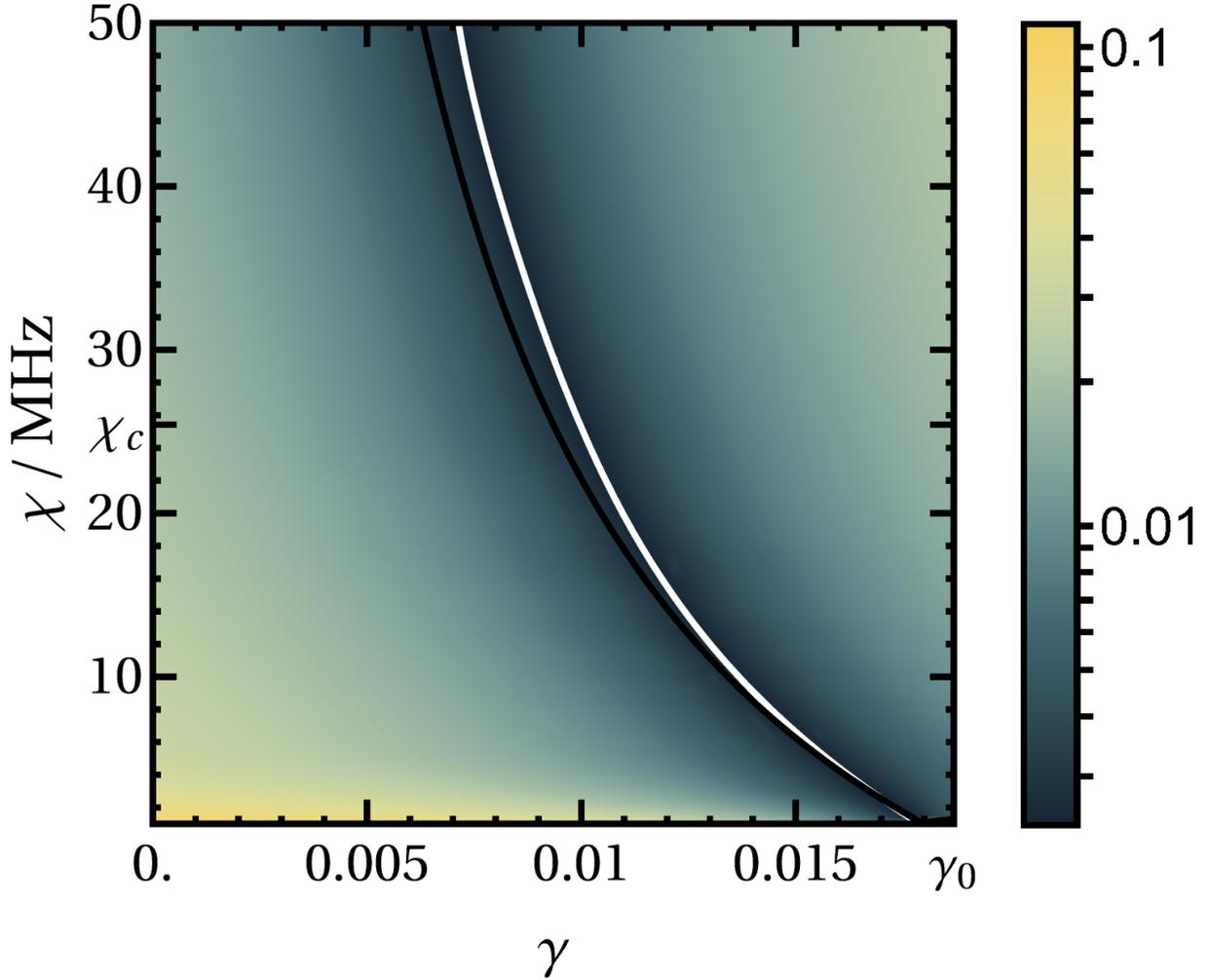}
 \caption{\label{fig3}$||\mathcal{E}-\mathcal{E}^i||_\diamond$ for a wide range of $\chi$. $\gamma$ characterises the rotation of the basis on which $\mathcal{E}^i$ projects compared to the bare basis, i.e. the left and right boundary of the figure correspond to the bare and dressed basis as seen in Fig.~\ref{fig2}. The numerically determined minimum is marked in white, while the black curve marks the $n=\alpha^2$-basis, Eq.~(\ref{na2-basis}).}
 \end{figure}

\begin{figure}
 \includegraphics[width=\columnwidth]{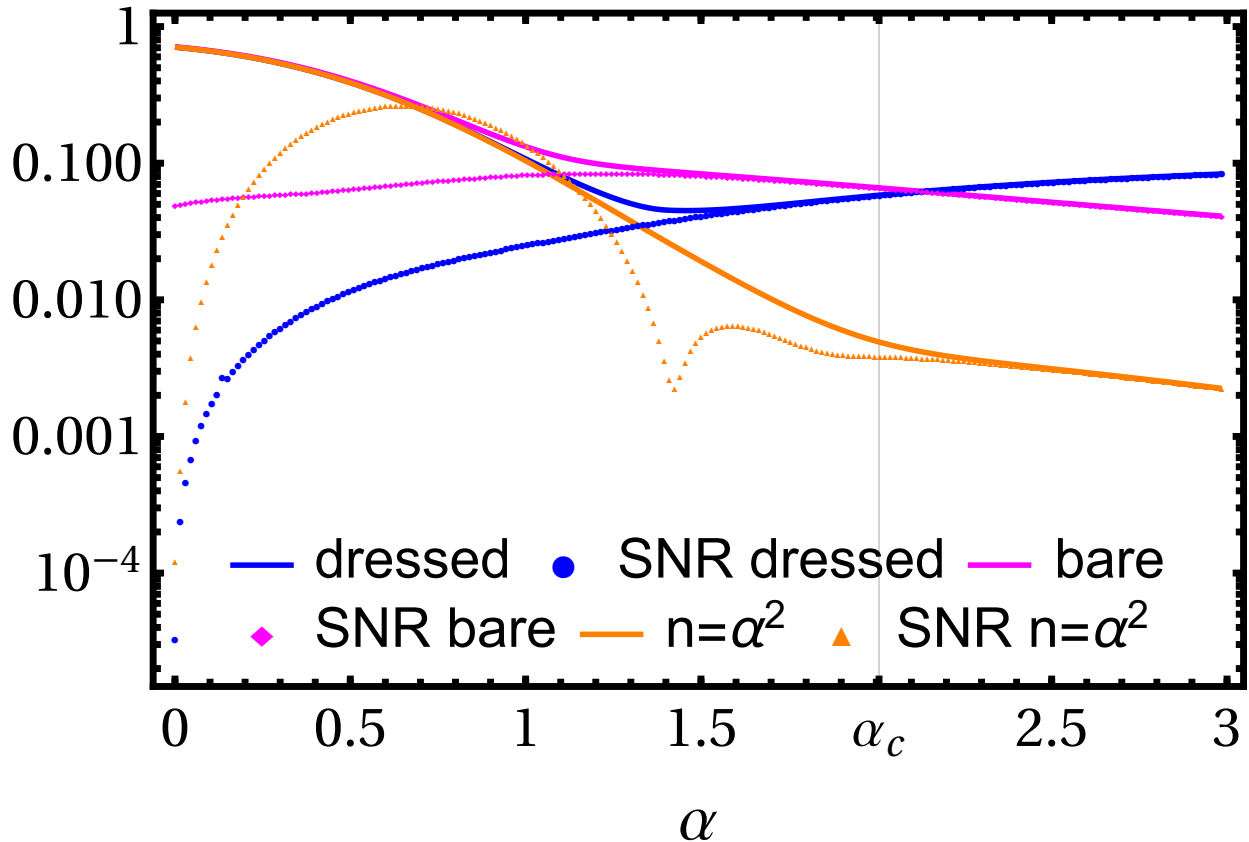}
 \caption{\label{fig4}$||\mathcal{E}-\mathcal{E}^i||_\diamond$ as a function of coherent probe state amplitude $\alpha$. In analogy to $\chi_c$, the predicted crossover-point from dressed to bare basis behaviour is $\alpha_c^2=\sqrt{\delta_0^2+J^2}/\chi$. The parameters, $\delta_0=80\,$MHz, $J=10\,$MHz, $\chi=20\,$MHz, are chosen to include $\alpha_c$ in the plot range while keeping $\alpha$ small enough that $n_{\text{max}}=40$ is sufficient.}
 \end{figure}

As predicted, the $n=\alpha^2$-basis, Eq.~(\ref{na2-basis}), achieves a better agreement than both bare and dressed basis for small $|\chi|$. For positive $\chi$ this curve is consistently close to the absolute minimum, but begins to deviate more from it for greater $\chi$, as one can see in Fig.~\ref{fig3}, where Fig.~\ref{fig2} is augmented with an additional $\gamma$-axis.
For faster measurement, $\chi\to\infty$, we observe how the minimum in Fig.~\ref{fig3} moves towards the bare basis ($\gamma=0$). The same happens if $\alpha$ is increased, which is also shown in Fig.~\ref{fig4}. This can also be understood by recalling that in the limit of $n\to\infty$, the $\gamma_n$-dressed basis $\ket{\psi_n^{1,..4}}$ becomes the bare basis. To summarise, the stronger we measure, the closer we get to measuring the bare basis.
 
Where the idealised $n=\alpha^2$ model already agrees very well with the Nigg-Girvin measurement, including the finite SNR can make the diamond distance another order of magnitude and more smaller, as shown in Fig.~\ref{fig2}. In this parameter regime the non-interacting model, Eq.~(\ref{chimatrix})~and~(\ref{na2-basis}), provides an extremely accurate, analytic description of the Nigg-Girvin measurement. The plot in Fig.~\ref{fig2} unfortunately also showcases some of the shortcomings of the diamond norm implementation, as it approaches its precision limits for very small values of the diamond norm. In this case, we can use the Choi-Jamio{\l}kowski norm instead.

At $-\chi_c$, $\gamma_{n=\alpha^2}\to\pm\pi/4$ becomes maximal which clearly does not match the reality of the Nigg-Girvin model, so there the $n=\alpha^2$ model fails, as both the small $|\chi|$ and $|J|\ll |\delta_n|(=0)$ approximation do not hold anymore. Generally, when $\chi$ has a different sign from $\delta_0$, the effective detuning $|\delta_n|$ becomes smaller for increasing $\chi n$ before it increases again, which leads to undesirable interaction between qubits, negatively impacting the measurement. A related consequence is that the range of $\gamma_n$ corresponding to positive $n$ becomes much larger (for typical parameters), changing from the interval $I_0$ between $0$ and $\gamma_0$ to $[-\pi/4,\pi/4]\setminus I_0$. This makes it harder to scan numerically.

One aspect of this particular measurement model is always fast, and this is the instantaneous initialisation. By loading the readout resonator with a coherent state instead of slowly populating it over time, one could argue that we are abruptly changing the basis. Indeed if we could change the basis adiabatically, we should be able to measure the occupation of the computational 
basis states. There are a few points to consider though: First in our model, if we have infinite time on our hands we can just take $\chi\to0$ for the same effect. Second, if we want to keep using the final state after measurement, we also need to change it back adiabatically. This should inform the pulses to use. Of course, population and depopulation of the readout resonator are tied to the same timescale $1/\kappa$ which brings us to the third point: For optimal readout, we usually want to use $1/\kappa\le1/2\chi$ \cite{Wallraff17} and not slower. 

Above $\alpha$ was chosen small enough to make evaluation of the sums simple ($n_{\text{max}}$ not too big), and large enough that it does not induce significant measurement errors on its own that are independent of the simulation error due to the RWA. Conceptually there is no reason for this restriction, as we can just as well include the error induced by a finite $\alpha$ in our idealised non-interacting measurement models, see Eq.~(\ref{chimatrix}). One can see how naturally these perform much better than their more idealised counterparts for smaller $\alpha$ before quickly approaching the $\alpha\to\infty$ limit in Fig.~\ref{fig4} which shows the $\alpha$-dependence of the relevant diamond norms.

\section{Conclusion}
We have presented a simple simulation of single-qubit measurements that accounts for the effect of two-qubit coupling. It is based on a discretised version of the standard continuous dispersive readout in which the stochastic is replaced by unitary evolution, thus making the model more accessible to concrete calculations.
We have argued why a slow measurement tends to project onto the dressed basis while a fast and/or strong measurement projects onto the bare basis, and shown that this intuitive picture holds true for the investigated model. However we have also found an optimal intermediate qubit basis that provides an even more accurate description for realistic parameters.

The disparity between gate and measurement basis this reveals can be traced back to the interaction that enables the measurement in the first place. A qubit state dependent shift of the resonator frequency allows us to indirectly measure the qubit but it also implies that as the readout resonator is populated with photons the qubit frequencies and thus their eigenbasis (or rotating frame) changes.
This mismatch between bases adds to error rates. The error could be on the order of $1\%$, if we e.g. read off the dressed basis error in Fig.~\ref{fig2}, making it a relevant concern for the quantum computer's error budget, not to mention that this also implies that localised errors on the physical qubits become correlated in a dressed basis.
These effects all increase the measurement error which would have to be counteracted by quantum error correction.

Alternatively, we could ask how the knowledge of the measurement basis can be used to mitigate its effect on measurement errors, apart from adding it to a growing list of quantum control constraints. At present we cannot offer an immediately helpful scheme to achieve this. While this information allows for a suitable rotation that changes between gate- and measurement basis to be applied before and after measurements, the added two-qubit gates would only increase the error in any near-term scenarios, and neither would this strategy scale to larger networks.

Future work will investigate how our results and those obtained within a homodyne-readout model apply to larger networks of qubits. In that case the qubit Hamiltonian is not analytically diagonalizable anymore, and perturbative methods are fraught with the dangers of frequency collisions. The dressing, while mostly nearest-neighbour, becomes slightly delocalized. This makes obtaining general results more challenging.

\begin{acknowledgments}
\section{Acknowledgments}
We thank the OpenSuperQ project (820363) of the EU Flagship on Quantum Technology, H2020-FETFLAG-2018-03, for support.
\end{acknowledgments}

\bibliography{20jan}

\begin{thebibliography}{20}%
\makeatletter
\providecommand \@ifxundefined [1]{%
 \@ifx{#1\undefined}
}%
\providecommand \@ifnum [1]{%
 \ifnum #1\expandafter \@firstoftwo
 \else \expandafter \@secondoftwo
 \fi
}%
\providecommand \@ifx [1]{%
 \ifx #1\expandafter \@firstoftwo
 \else \expandafter \@secondoftwo
 \fi
}%
\providecommand \natexlab [1]{#1}%
\providecommand \enquote  [1]{``#1''}%
\providecommand \bibnamefont  [1]{#1}%
\providecommand \bibfnamefont [1]{#1}%
\providecommand \citenamefont [1]{#1}%
\providecommand \href@noop [0]{\@secondoftwo}%
\providecommand \href [0]{\begingroup \@sanitize@url \@href}%
\providecommand \@href[1]{\@@startlink{#1}\@@href}%
\providecommand \@@href[1]{\endgroup#1\@@endlink}%
\providecommand \@sanitize@url [0]{\catcode `\\12\catcode `\$12\catcode
  `\&12\catcode `\#12\catcode `\^12\catcode `\_12\catcode `\%12\relax}%
\providecommand \@@startlink[1]{}%
\providecommand \@@endlink[0]{}%
\providecommand \url  [0]{\begingroup\@sanitize@url \@url }%
\providecommand \@url [1]{\endgroup\@href {#1}{\urlprefix }}%
\providecommand \urlprefix  [0]{URL }%
\providecommand \Eprint [0]{\href }%
\providecommand \doibase [0]{https://doi.org/}%
\providecommand \selectlanguage [0]{\@gobble}%
\providecommand \bibinfo  [0]{\@secondoftwo}%
\providecommand \bibfield  [0]{\@secondoftwo}%
\providecommand \translation [1]{[#1]}%
\providecommand \BibitemOpen [0]{}%
\providecommand \bibitemStop [0]{}%
\providecommand \bibitemNoStop [0]{.\EOS\space}%
\providecommand \EOS [0]{\spacefactor3000\relax}%
\providecommand \BibitemShut  [1]{\csname bibitem#1\endcsname}%
\let\auto@bib@innerbib\@empty
\bibitem [{\citenamefont {Blais}\ \emph {et~al.}(2004)\citenamefont {Blais},
  \citenamefont {Huang}, \citenamefont {Wallraff}, \citenamefont {Girvin},\
  and\ \citenamefont {Schoelkopf}}]{Blais04}%
  \BibitemOpen
  \bibfield  {author} {\bibinfo {author} {\bibfnamefont {A.}~\bibnamefont
  {Blais}}, \bibinfo {author} {\bibfnamefont {R.-S.}\ \bibnamefont {Huang}},
  \bibinfo {author} {\bibfnamefont {A.}~\bibnamefont {Wallraff}}, \bibinfo
  {author} {\bibfnamefont {S.~M.}\ \bibnamefont {Girvin}},\ and\ \bibinfo
  {author} {\bibfnamefont {R.~J.}\ \bibnamefont {Schoelkopf}},\ }\bibfield
  {title} {\bibinfo {title} {Cavity quantum electrodynamics for superconducting
  electrical circuits: An architecture for quantum computation},\ }\href
  {https://doi.org/10.1103/PhysRevA.69.062320} {\bibfield  {journal} {\bibinfo
  {journal} {Phys. Rev. A}\ }\textbf {\bibinfo {volume} {69}},\ \bibinfo
  {pages} {062320} (\bibinfo {year} {2004})}\BibitemShut {NoStop}%
\bibitem [{\citenamefont {Walter}\ \emph {et~al.}(2017)\citenamefont {Walter},
  \citenamefont {Kurpiers}, \citenamefont {Gasparinetti}, \citenamefont
  {Magnard}, \citenamefont {{Poto\ifmmode \check{c}\else {\v c}\fi{}nik}},
  \citenamefont {Salath{\'e}}, \citenamefont {Pechal}, \citenamefont {Mondal},
  \citenamefont {Oppliger}, \citenamefont {Eichler},\ and\ \citenamefont
  {Wallraff}}]{Wallraff17}%
  \BibitemOpen
  \bibfield  {author} {\bibinfo {author} {\bibfnamefont {T.}~\bibnamefont
  {Walter}}, \bibinfo {author} {\bibfnamefont {P.}~\bibnamefont {Kurpiers}},
  \bibinfo {author} {\bibfnamefont {S.}~\bibnamefont {Gasparinetti}}, \bibinfo
  {author} {\bibfnamefont {P.}~\bibnamefont {Magnard}}, \bibinfo {author}
  {\bibfnamefont {A.}~\bibnamefont {{Poto\ifmmode \check{c}\else {\v
  c}\fi{}nik}}}, \bibinfo {author} {\bibfnamefont {Y.}~\bibnamefont
  {Salath{\'e}}}, \bibinfo {author} {\bibfnamefont {M.}~\bibnamefont {Pechal}},
  \bibinfo {author} {\bibfnamefont {M.}~\bibnamefont {Mondal}}, \bibinfo
  {author} {\bibfnamefont {M.}~\bibnamefont {Oppliger}}, \bibinfo {author}
  {\bibfnamefont {C.}~\bibnamefont {Eichler}},\ and\ \bibinfo {author}
  {\bibfnamefont {A.}~\bibnamefont {Wallraff}},\ }\bibfield  {title} {\bibinfo
  {title} {Rapid high-fidelity single-shot dispersive readout of
  superconducting qubits},\ }\href
  {https://doi.org/10.1103/PhysRevApplied.7.054020} {\bibfield  {journal}
  {\bibinfo  {journal} {Phys. Rev. Applied}\ }\textbf {\bibinfo {volume} {7}},\
  \bibinfo {pages} {054020} (\bibinfo {year} {2017})}\BibitemShut {NoStop}%
\bibitem [{\citenamefont {Heinsoo}\ \emph {et~al.}(2018)\citenamefont
  {Heinsoo}, \citenamefont {Andersen}, \citenamefont {Remm}, \citenamefont
  {Krinner}, \citenamefont {Walter}, \citenamefont {Salath{\'e}}, \citenamefont
  {Gasparinetti}, \citenamefont {Besse}, \citenamefont {{Poto\ifmmode
  \check{c}\else {\v c}\fi{}nik}}, \citenamefont {Wallraff},\ and\
  \citenamefont {Eichler}}]{Wallraff18}%
  \BibitemOpen
  \bibfield  {author} {\bibinfo {author} {\bibfnamefont {J.}~\bibnamefont
  {Heinsoo}}, \bibinfo {author} {\bibfnamefont {C.~K.}\ \bibnamefont
  {Andersen}}, \bibinfo {author} {\bibfnamefont {A.}~\bibnamefont {Remm}},
  \bibinfo {author} {\bibfnamefont {S.}~\bibnamefont {Krinner}}, \bibinfo
  {author} {\bibfnamefont {T.}~\bibnamefont {Walter}}, \bibinfo {author}
  {\bibfnamefont {Y.}~\bibnamefont {Salath{\'e}}}, \bibinfo {author}
  {\bibfnamefont {S.}~\bibnamefont {Gasparinetti}}, \bibinfo {author}
  {\bibfnamefont {J.-C.}\ \bibnamefont {Besse}}, \bibinfo {author}
  {\bibfnamefont {A.}~\bibnamefont {{Poto\ifmmode \check{c}\else {\v
  c}\fi{}nik}}}, \bibinfo {author} {\bibfnamefont {A.}~\bibnamefont
  {Wallraff}},\ and\ \bibinfo {author} {\bibfnamefont {C.}~\bibnamefont
  {Eichler}},\ }\bibfield  {title} {\bibinfo {title} {Rapid high-fidelity
  multiplexed readout of superconducting qubits},\ }\href
  {https://doi.org/10.1103/PhysRevApplied.10.034040} {\bibfield  {journal}
  {\bibinfo  {journal} {Phys. Rev. Applied}\ }\textbf {\bibinfo {volume}
  {10}},\ \bibinfo {pages} {034040} (\bibinfo {year} {2018})}\BibitemShut
  {NoStop}%
\bibitem [{\citenamefont {Wendin}(2017)}]{Wendin17}%
  \BibitemOpen
  \bibfield  {author} {\bibinfo {author} {\bibfnamefont {G.}~\bibnamefont
  {Wendin}},\ }\bibfield  {title} {\bibinfo {title} {Quantum information
  processing with superconducting circuits: a review},\ }\href
  {https://doi.org/10.1088/1361-6633/aa7e1a} {\bibfield  {journal} {\bibinfo
  {journal} {Rep. Prog. Phys.}\ }\textbf {\bibinfo {volume} {80}},\ \bibinfo
  {pages} {106001} (\bibinfo {year} {2017})}\BibitemShut {NoStop}%
\bibitem [{\citenamefont {Khezri}\ \emph {et~al.}(2015)\citenamefont {Khezri},
  \citenamefont {Dressel},\ and\ \citenamefont {Korotkov}}]{Khezri15}%
  \BibitemOpen
  \bibfield  {author} {\bibinfo {author} {\bibfnamefont {M.}~\bibnamefont
  {Khezri}}, \bibinfo {author} {\bibfnamefont {J.}~\bibnamefont {Dressel}},\
  and\ \bibinfo {author} {\bibfnamefont {A.~N.}\ \bibnamefont {Korotkov}},\
  }\bibfield  {title} {\bibinfo {title} {Qubit measurement error from coupling
  with a detuned neighbor in circuit {QED}},\ }\href
  {https://doi.org/10.1103/PhysRevA.92.052306} {\bibfield  {journal} {\bibinfo
  {journal} {Phys. Rev. A}\ }\textbf {\bibinfo {volume} {92}},\ \bibinfo
  {pages} {052306} (\bibinfo {year} {2015})}\BibitemShut {NoStop}%
\bibitem [{\citenamefont {Ashhab}\ \emph
  {et~al.}(2009{\natexlab{a}})\citenamefont {Ashhab}, \citenamefont {You},\
  and\ \citenamefont {Nori}}]{Ashhab09PRA}%
  \BibitemOpen
  \bibfield  {author} {\bibinfo {author} {\bibfnamefont {S.}~\bibnamefont
  {Ashhab}}, \bibinfo {author} {\bibfnamefont {J.~Q.}\ \bibnamefont {You}},\
  and\ \bibinfo {author} {\bibfnamefont {F.}~\bibnamefont {Nori}},\ }\bibfield
  {title} {\bibinfo {title} {Weak and strong measurement of a qubit using a
  switching-based detector},\ }\href
  {https://doi.org/10.1103/PhysRevA.79.032317} {\bibfield  {journal} {\bibinfo
  {journal} {Phys. Rev. A}\ }\textbf {\bibinfo {volume} {79}},\ \bibinfo
  {pages} {032317} (\bibinfo {year} {2009}{\natexlab{a}})}\BibitemShut
  {NoStop}%
\bibitem [{\citenamefont {Ashhab}\ \emph
  {et~al.}(2009{\natexlab{b}})\citenamefont {Ashhab}, \citenamefont {You},\
  and\ \citenamefont {Nori}}]{Ashhab09NJP}%
  \BibitemOpen
  \bibfield  {author} {\bibinfo {author} {\bibfnamefont {S.}~\bibnamefont
  {Ashhab}}, \bibinfo {author} {\bibfnamefont {J.~Q.}\ \bibnamefont {You}},\
  and\ \bibinfo {author} {\bibfnamefont {F.}~\bibnamefont {Nori}},\ }\bibfield
  {title} {\bibinfo {title} {The information about the state of a qubit gained
  by a weakly coupled detector},\ }\href
  {https://doi.org/10.1088/1367-2630/11/8/083017} {\bibfield  {journal}
  {\bibinfo  {journal} {New Journal of Physics}\ }\textbf {\bibinfo {volume}
  {11}},\ \bibinfo {pages} {083017} (\bibinfo {year}
  {2009}{\natexlab{b}})}\BibitemShut {NoStop}%
\bibitem [{\citenamefont {de~Groot}\ \emph {et~al.}(2010)\citenamefont
  {de~Groot}, \citenamefont {Lisenfeld}, \citenamefont {Schouten},
  \citenamefont {Ashhab}, \citenamefont {Lupa{\c s}cu}, \citenamefont
  {Harmans},\ and\ \citenamefont {Mooij}}]{deGroot10}%
  \BibitemOpen
  \bibfield  {author} {\bibinfo {author} {\bibfnamefont {P.~C.}\ \bibnamefont
  {de~Groot}}, \bibinfo {author} {\bibfnamefont {J.}~\bibnamefont {Lisenfeld}},
  \bibinfo {author} {\bibfnamefont {R.~N.}\ \bibnamefont {Schouten}}, \bibinfo
  {author} {\bibfnamefont {S.}~\bibnamefont {Ashhab}}, \bibinfo {author}
  {\bibfnamefont {A.}~\bibnamefont {Lupa{\c s}cu}}, \bibinfo {author}
  {\bibfnamefont {C.~J. P.~M.}\ \bibnamefont {Harmans}},\ and\ \bibinfo
  {author} {\bibfnamefont {J.~E.}\ \bibnamefont {Mooij}},\ }\bibfield  {title}
  {\bibinfo {title} {Selective darkening of degenerate transitions demonstrated
  with two superconducting quantum bits},\ }\href
  {https://doi.org/10.1038/nphys1733} {\bibfield  {journal} {\bibinfo
  {journal} {Nature Physics}\ }\textbf {\bibinfo {volume} {6}},\ \bibinfo
  {pages} {763} (\bibinfo {year} {2010})}\BibitemShut {NoStop}%
\bibitem [{\citenamefont {Gambetta}\ \emph {et~al.}(2017)\citenamefont
  {Gambetta}, \citenamefont {Chow},\ and\ \citenamefont
  {Steffen}}]{Gambetta17}%
  \BibitemOpen
  \bibfield  {author} {\bibinfo {author} {\bibfnamefont {J.~M.}\ \bibnamefont
  {Gambetta}}, \bibinfo {author} {\bibfnamefont {J.~M.}\ \bibnamefont {Chow}},\
  and\ \bibinfo {author} {\bibfnamefont {M.}~\bibnamefont {Steffen}},\
  }\bibfield  {title} {\bibinfo {title} {Building logical qubits in a
  superconducting quantum computing system},\ }\href
  {https://doi.org/10.1038/s41534-016-0004-0} {\bibfield  {journal} {\bibinfo
  {journal} {npj Quantum Information}\ }\textbf {\bibinfo {volume} {3}},\
  \bibinfo {pages} {2} (\bibinfo {year} {2017})}\BibitemShut {NoStop}%
\bibitem [{\citenamefont {Nigg}\ and\ \citenamefont {Girvin}(2013)}]{Nigg13}%
  \BibitemOpen
  \bibfield  {author} {\bibinfo {author} {\bibfnamefont {S.~E.}\ \bibnamefont
  {Nigg}}\ and\ \bibinfo {author} {\bibfnamefont {S.~M.}\ \bibnamefont
  {Girvin}},\ }\bibfield  {title} {\bibinfo {title} {Stabilizer quantum error
  correction toolbox for superconducting qubits},\ }\href
  {https://doi.org/10.1103/PhysRevLett.110.243604} {\bibfield  {journal}
  {\bibinfo  {journal} {Phys. Rev. Lett.}\ }\textbf {\bibinfo {volume} {110}},\
  \bibinfo {pages} {243604} (\bibinfo {year} {2013})}\BibitemShut {NoStop}%
\bibitem [{\citenamefont {Richer}(2013)}]{Richer13}%
  \BibitemOpen
  \bibfield  {author} {\bibinfo {author} {\bibfnamefont {S.}~\bibnamefont
  {Richer}},\ }\emph {\bibinfo {title} {Perturbative analysis of two-qubit
  gates on transmon qubits}},\ \href
  {https://www.quantuminfo.physik.rwth-aachen.de/global/show_document.asp?id=aaaaaaaaaajiobd}
  {Master's thesis},\ \bibinfo  {school} {RWTH Aachen University} (\bibinfo
  {year} {2013})\BibitemShut {NoStop}%
\bibitem [{\citenamefont {Gambetta}(2013)}]{Gambetta13}%
  \BibitemOpen
  \bibfield  {author} {\bibinfo {author} {\bibfnamefont {J.~M.}\ \bibnamefont
  {Gambetta}},\ }\bibfield  {title} {\bibinfo {title} {Control of
  superconducting qubits},\ }in\ \href@noop {} {\emph {\bibinfo {booktitle}
  {Lecture Notes of the 44th IFF Spring School, Quantum Information
  Processing}}},\ \bibinfo {series} {Schl{\"u}sseltechnologien/Key
  Technologies}, Vol.~\bibinfo {volume} {52},\ \bibinfo {editor} {edited by\
  \bibinfo {editor} {\bibfnamefont {D.}~\bibnamefont {DiVincenzo}}}\ (\bibinfo
  {publisher} {Schriften des Forschungszentrums J{\"u}lich},\ \bibinfo {year}
  {2013})\BibitemShut {NoStop}%
\bibitem [{\citenamefont {Magesan}\ and\ \citenamefont
  {Gambetta}(2020)}]{Magesan20}%
  \BibitemOpen
  \bibfield  {author} {\bibinfo {author} {\bibfnamefont {E.}~\bibnamefont
  {Magesan}}\ and\ \bibinfo {author} {\bibfnamefont {J.~M.}\ \bibnamefont
  {Gambetta}},\ }\bibfield  {title} {\bibinfo {title} {Effective {H}amiltonian
  models of the cross-resonance gate},\ }\href
  {https://doi.org/10.1103/PhysRevA.101.052308} {\bibfield  {journal} {\bibinfo
   {journal} {Phys. Rev. A}\ }\textbf {\bibinfo {volume} {101}},\ \bibinfo
  {pages} {052308} (\bibinfo {year} {2020})}\BibitemShut {NoStop}%
\bibitem [{\citenamefont {Peres}(1993)}]{Peres93}%
  \BibitemOpen
  \bibfield  {author} {\bibinfo {author} {\bibfnamefont {A.}~\bibnamefont
  {Peres}},\ }\href {https://books.google.de/books?id=-yLkUyCsb7cC} {\emph
  {\bibinfo {title} {Quantum Theory: Concepts and Methods}}},\ Fundamental
  Theories of Physics\ (\bibinfo  {publisher} {Springer Netherlands},\ \bibinfo
  {year} {1993})\BibitemShut {NoStop}%
\bibitem [{\citenamefont {Gambetta}\ \emph {et~al.}(2007)\citenamefont
  {Gambetta}, \citenamefont {Braff}, \citenamefont {Wallraff}, \citenamefont
  {Girvin},\ and\ \citenamefont {Schoelkopf}}]{Gambetta07}%
  \BibitemOpen
  \bibfield  {author} {\bibinfo {author} {\bibfnamefont {J.}~\bibnamefont
  {Gambetta}}, \bibinfo {author} {\bibfnamefont {W.~A.}\ \bibnamefont {Braff}},
  \bibinfo {author} {\bibfnamefont {A.}~\bibnamefont {Wallraff}}, \bibinfo
  {author} {\bibfnamefont {S.~M.}\ \bibnamefont {Girvin}},\ and\ \bibinfo
  {author} {\bibfnamefont {R.~J.}\ \bibnamefont {Schoelkopf}},\ }\bibfield
  {title} {\bibinfo {title} {Protocols for optimal readout of qubits using a
  continuous quantum nondemolition measurement},\ }\href
  {https://doi.org/10.1103/PhysRevA.76.012325} {\bibfield  {journal} {\bibinfo
  {journal} {Phys. Rev. A}\ }\textbf {\bibinfo {volume} {76}},\ \bibinfo
  {pages} {012325} (\bibinfo {year} {2007})}\BibitemShut {NoStop}%
\bibitem [{\citenamefont {Galiautdinov}\ \emph {et~al.}(2012)\citenamefont
  {Galiautdinov}, \citenamefont {Korotkov},\ and\ \citenamefont
  {Martinis}}]{Galiautdinov12}%
  \BibitemOpen
  \bibfield  {author} {\bibinfo {author} {\bibfnamefont {A.}~\bibnamefont
  {Galiautdinov}}, \bibinfo {author} {\bibfnamefont {A.~N.}\ \bibnamefont
  {Korotkov}},\ and\ \bibinfo {author} {\bibfnamefont {J.~M.}\ \bibnamefont
  {Martinis}},\ }\bibfield  {title} {\bibinfo {title} {Resonator--zero-qubit
  architecture for superconducting qubits},\ }\href
  {https://doi.org/10.1103/PhysRevA.85.042321} {\bibfield  {journal} {\bibinfo
  {journal} {Phys. Rev. A}\ }\textbf {\bibinfo {volume} {85}},\ \bibinfo
  {pages} {042321} (\bibinfo {year} {2012})}\BibitemShut {NoStop}%
\bibitem [{\citenamefont {Aharonov}\ \emph {et~al.}(1998)\citenamefont
  {Aharonov}, \citenamefont {Kitaev},\ and\ \citenamefont {Nisan}}]{Kitaev98}%
  \BibitemOpen
  \bibfield  {author} {\bibinfo {author} {\bibfnamefont {D.}~\bibnamefont
  {Aharonov}}, \bibinfo {author} {\bibfnamefont {A.}~\bibnamefont {Kitaev}},\
  and\ \bibinfo {author} {\bibfnamefont {N.}~\bibnamefont {Nisan}},\ }\bibfield
   {title} {\bibinfo {title} {Quantum circuits with mixed states},\ }\bibfield
  {journal} {\bibinfo  {journal} {Proceedings of the thirtieth annual ACM
  symposium on Theory of computing - STOC {\rq}98}\ }\href
  {https://doi.org/10.1145/276698.276708} {10.1145/276698.276708} (\bibinfo
  {year} {1998})\BibitemShut {NoStop}%
\bibitem [{\citenamefont {Watrous}(2009)}]{Watrous09}%
  \BibitemOpen
  \bibfield  {author} {\bibinfo {author} {\bibfnamefont {J.}~\bibnamefont
  {Watrous}},\ }\bibfield  {title} {\bibinfo {title} {Semidefinite programs for
  completely bounded norms},\ }\href
  {https://doi.org/10.4086/toc.2009.v005a011} {\bibfield  {journal} {\bibinfo
  {journal} {Theory of Computing}\ }\textbf {\bibinfo {volume} {5}},\ \bibinfo
  {pages} {217} (\bibinfo {year} {2009})}\BibitemShut {NoStop}%
\bibitem [{\citenamefont {Watrous}(2013)}]{Watrous12}%
  \BibitemOpen
  \bibfield  {author} {\bibinfo {author} {\bibfnamefont {J.}~\bibnamefont
  {Watrous}},\ }\bibfield  {title} {\bibinfo {title} {Simpler semidefinite
  programs for completely bounded norms},\ }\bibfield  {journal} {\bibinfo
  {journal} {Chicago Journal of Theoretical Computer Science}\ }\textbf
  {\bibinfo {volume} {2013}},\ \href {https://doi.org/10.4086/cjtcs.2013.008}
  {10.4086/cjtcs.2013.008} (\bibinfo {year} {2013})\BibitemShut {NoStop}%
\bibitem [{\citenamefont {Watrous}(2018)}]{Watrous18}%
  \BibitemOpen
  \bibfield  {author} {\bibinfo {author} {\bibfnamefont {J.}~\bibnamefont
  {Watrous}},\ }\href
  {https://www.cambridge.org/de/academic/subjects/computer-science/algorithmics-complexity-computer-algebra-and-computational-g/theory-quantum-information?format=HB&isbn=9781107180567}
  {\emph {\bibinfo {title} {The Theory of Quantum Information}}}\ (\bibinfo
  {publisher} {Cambridge University Press},\ \bibinfo {year}
  {2018})\BibitemShut {NoStop}%
\end{thebibliography}%

\end{document}